\documentclass[prl,twocolumn,aps,amsmath,amssymb]{revtex4}

\usepackage{graphicx}
\usepackage{dcolumn}
\usepackage{bm}
\usepackage{wrapfig}
\usepackage{amsmath,amsfonts,amssymb,bm,ulem}
\usepackage[usenames]{color}
\usepackage{bm}
\usepackage{slashed}
\usepackage{mathrsfs}
\usepackage{graphicx}
\usepackage{dcolumn}
\usepackage{bm}

\usepackage{hyperref}
\usepackage{cleveref}

\newcommand{\be}{\begin{equation}}
\newcommand{\ee}{\end{equation}}

\begin{document}

\title{Yang-Mills Gauss law and the heavy quark binding energy in the presence of a dimension-2 gluon condensate}

\author{Jordan Wilson-Gerow}
\affiliation{Department of Physics and Astronomy, University
	of British Columbia, 6224 Agricultural Rd., Vancouver, B.C., Canada
	V6T 1Z1}
\date{\today}

\begin{abstract}
We study the binding energy of a heavy quark-antiquark ($q\bar{q}$) pair using the first-order path integral formalism.  This makes the Yang-Mills constraint equation explicit, and highlights that it is valid without relying on a semiclassical approximation. A generalized gauge-covariant Coulomb gauge is chosen to allow for a decomposition of the chromoelectric field into a gauge-covariant generalization of transverse and longitudinal parts. This decomposition makes it clear that the  $q\bar{q}$ binding energy is determined solely by the solution to the constraint equation. Assuming that the low-energy physics is dominated by the existence of a dimension-2 gluon condensate, we develop an asymptotic series solution to the constraint equation and thus to the $q\bar{q}$ binding energy. We predict a QCD string tension in terms of the condensate strength and quadratic Casimir eigenvalues, and relate our result to results coming from OPE analyses.
\end{abstract}

\maketitle


\section{I: Introduction}
\label{sec:intro}

The discovery of asymptotic freedom in Yang-Mills theory signaled a blessing and a curse for the theory~\cite{asyfree1973}. The discovery demonstrated that perturbative calculations could be reliable and experimentally testable at high-energies, allowing for high-energy experiments to confirm that $SU(3)$ Yang-Mills theory is the correct fundamental theory of the strong nuclear interaction.  Simultaneously though, it suggested that low energy phenomenon could not be described using weak coupling methods, such as perturbation theory or semi-classical approximations. 

Perhaps the most important low energy phenomenon to understand in Yang-Mills theory is hadron formation. At low energies one never observes individual quarks; rather, these objects with non-trivial color charge are always bound together in colorless hadronic states. Despite having this empirical observation for over 55 years~\cite{gellmann1964}, we still do not have an accepted theoretical post-diction of the phenomenon, commonly referred to as quark (or color-) confinement.

Early analysis of the various mesonic states observed in nature found that they quite accurately fit into linear Regge trajectories~\cite{regge1959,chew196162}, a hallmark feature of the spectrum of a relativistic quantum string~\cite{nambu1969,nielsen1970,susskind1970,chang1972,goddard1973}. Furthermore, one could read off the effective ``QCD-string'' tension from these trajectories. This observation narrowed the study for quark confinement. One then had the idea the some underlying low-energy Yang-Mills physics should lead to the emergence of an effective string of chromoelectric flux connecting two quarks~\cite{kogut1974}. It remained, however, to understand why and how the chromoelectric field lines in non-abelian Yang-Mills theory would bunch together to form a string-like object rather than spreading radially as they would in electrodynamics. A variety of proposals have been given for this mechanism, many of which involving a non-trivial vacuum structure arising from non-perturbative topological excitations such as magnetic monopoles, instantons, dyons, and center vortices, see for example~\cite{mandelstam1976,thooft1975,thooft1979,nielsen1979,cornwall1979,ambjorn1980,mack1980,thooft1981,mack1982,debbio1998,chernodub1999}.

The most useful tool we've had to understand low-energy Yang-Mills physics has been the lattice simulations. By discretizing the underlying spacetime one simultaneously provides a necessary short-distance regularization of the theory and also truncates to a finite number of degrees of freedom so that the the theory can be simulated on a computer. This approach has been remarkably successful (see some of the many reviews for a summary~\cite{latticereview}), the result most relevent to us being the confirmation of the above intuition that the interaction potential between a quark and anti-quark is Coulombic at short distances where perturbation theory is applicable, but becomes linear at larger distances (see eg.~\cite{bali1992,bali1993,bali1997,bali2000c,bali2000d,bolder2001,necco2001}).

Despite its successes, lattice simulations have some limitations as theoretical tools. Firstly, to ensure proper convergence of the path-integral one is typically restricted to a Euclidean rather than Minkowskian description. This considerably limits the types of observables one can compute, because it obfuscates time dependence. Secondly, and more importantly, the simulations primarily function as a ``black-box''.  One provides the YM action and coupling constant, and a lattice structure, and the computer code will compute observables. From simulations we do not necessarily get an intuitive physical description of the QCD string formation phenomenon, nor do we necessarily get clear insights on how to better predict the phenomenon analytically. 

Recently though, there has been interest in the lattice community in studying the effects of a gluon condensate~\cite{boucaud2000a,boucaud2000b,destoto2001,boucaud2001,boucaud2002}. Notably, one finds that lattice measurements of the running coupling constant are not well fit by models unless the models include significant contributions from a dimension-2 gluon condensate, $\langle A^{a}_{\mu}A^{a}_{\mu}\rangle\neq0$.~\cite{boucaud2000b,destoto2001} Using operator product expansion (OPE) techniques, relationships have been made between this condensate and quark confinement. Additionally, using lattice simulations it has been observed that are relationships between this condensate and a vacuum described by a topological instanton liquid~\cite{boucaud2002}. This observation hints that if we more deeply understand the dimension-2 condensate, and the phenomenological consequences of its existence, we may come full circle back to the original topological/non-perturbative intuitions for the nature of quark confinement~\cite{mandelstam1976,thooft1975,thooft1981,nielsen1979}.

In this paper we aim to provide a new analytical approach to the study of the quark confinement problem. We use a heavy quark approximation to study a quark-antiquark pair at fixed separation. We use the path-integral, and assume that the vacuum is described by a gluon condensate. We do not remark however on the mechanism which generates this condensate. The approach we use here has two primary features. Firstly, we use the first-order path-integral formalism so that the chromoelectric field is an explicit variable in the path-integral. Yang-Mills theory is a constrained theory, and in the first order formalism one sees explicitly that the chromoelectric field must satisfy the Yang-Mills analog of the Gauss law. Since it is a constraint, this equation holds exactly and does not rely on a semiclassical approximation to be valid in the quantum theory. This point is essential, as we do not expect a semi-classical description to be valid as the coupling flows to larger values. Secondly, we use a judicious gauge choice as well as a generalization of the transverse and longitudinal fields familiar from electrodynamics.  These choices allow for a decomposition of the field variables such that the contribution to the quark-antiquark binding energy is determined solely by the solution to the constraint equation. We proceed to compute this binding energy as an asymptotic series with increasing powers of the particle separation $r$.

In section II we review the Yang-Mills path integral, introducing the: first-order formalism, gauge-covariant Coulomb gauge, and gauge-covariant transverse-longitudinal decomposition. We further demonstrate how with this choice of variables we can isolate the contributions to the quark-antiquark static potential as arising solely from the solution to the constraint equation. 

In section III we discuss the constraint equation and introduce a path-integral description of the condensate following that of~\cite{celenza1986}. We then develop an asymptotic series expansion in powers of $r$ for the interaction potential, with higher order terms predicted by a recursion relation.

 In section IV we relate our approach to the study of Wilson loops, and proceed to compute the first few terms in the interaction energy. We predict a Yang-Mills string tension in terms of the zero-energy values of the condensate strength, the zero-energy value of the coupling constant, and the quadratic Casimir eigenvalues of the adjoint representation and chosen quark representation. 
 
 In section V we relate our prediction to predictions coming from OPE analyses and comment on various deficiencies of our model.

\section{II: Non-abelian Yang Mills Theory Formalism}
\label{sec:naYM}

In Yang-Mills theory there is a gauge field $A_{\mu}$ which takes values in the Lie-algebra of the group $G$. From this one constructs the field strength tensor
\begin{equation}
F_{\mu\nu}=\partial_{\mu}A_{\nu}-\partial_{\nu}A_{\mu}-i[A_{\mu},A_{\nu}],
\end{equation}
and then the action
\begin{equation}\label{eq:YMaction}
S=\int d^{4}x\bigg(-\frac{1}{2g^{2}}\textrm{tr}\,F_{\mu\nu}F^{\mu\nu}+\textrm{tr}A_{\mu}J^{\mu}\bigg).
\end{equation}
Here the trace is taken in the fundamental (lowest dimension) representation of the gauge group, and the generators $T^{a}$, $a=1,...,\textrm{dim(G)}$, in the fundamental representation have been normalized as usual
\begin{equation}
\textrm{tr}\,T^{a}T^{b}=\frac{1}{2}\delta^{ab}.
\end{equation}
The source $J^{\mu}$ is a Lie-algebra valued current, and we'll specify to the case that it is the sum of contributions from particles on fixed worldlines
\begin{equation}
J^{\mu}(y)=\sum_{n}\int d\tau \frac{d x^{\mu}_{n}}{d\tau}\,q_{n}(\tau)\delta^{4}(y-x_{n}(\tau)).
\end{equation}
In this we've introduced the Lie-algebra valued, time dependent, \textit{color charge} $q_{n}(\tau)$.

Under infinitesimal gauge transformations $\Omega\approx 1+i\omega$, with $\omega$ in the Lie-algebra, we have
\begin{align}
\delta F_{\mu\nu}&=i[\omega,F_{\mu\nu}] \nonumber \\
\delta A_{\mu}&=\partial_{\mu}\omega-i[A_{\mu},\omega]\equiv D_{\mu}(A)\omega.
\end{align}
The gauge-covariant derivative $D_{\mu}(A)$ defined here will appear frequently throughout the following calculations. From these transformation rules we can see that the Yang-Mills action is gauge invariant if the current satisfies
\begin{equation}
D_{\mu}(A)J^{\mu}=0.
\end{equation}
Since $A$ is a dynamical quantum variable, this is not a requirement which could be imposed on a fixed external current, and so the internal \textit{color} degree of freedom must also be quantum mechanical. It remains consistent though to fix the particle worldlines. This is effectively an approximation in which the masses of the charges have been taken to be arbitrarily large. For the time being, we'll treat $J^{\mu}$ as fixed and then later introduce the internal dynamics for the color variables. 

The Yang-Mills generating functional is then just the path-integral
\begin{equation}\label{eq:YMgf}
Z[J]=\int\mathcal{D}A_{\mu}\,e^{i\int d^{4}x\big(-\frac{1}{2g^{2}}\textrm{tr}\,F_{\mu\nu}F^{\mu\nu}+\textrm{tr}A_{\mu}J^{\mu}\big)}.
\end{equation} 
We can transition to a first-order form  by introducing a Lie-algebra valued \textit{chromoelectric} field variable $E_{j}$ if we multiply~\cref{eq:YMgf} by $1$, represented as a gaussian integral.
 For convenience we also use the conventional redefinition $A_{\mu}\rightarrow gA_{\mu},\, F^{\mu\nu}\rightarrow\partial_{\mu}A_{\nu}-\partial_{\nu}A_{\mu}-ig[A_{\mu},A_{\nu}]$ to make the coupling constant appear with the interaction terms. The result is
 \begin{equation}
 Z[J]=\int\mathcal{D}E_{j}\,e^{-i\textrm{tr}\int d^{4}x\big(E_{j}+(\partial_{0}A_{j}-\partial_{j}A_{0})\big)^{2}}Z[J],
 \end{equation}
 which can be better rewritten as
\begin{equation}
Z[J]=\int\mathcal{D}E_{j}\int\mathcal{D}A_{\mu}\,e^{iS},
\end{equation}
with first-order action
\begin{align}\label{eq:foaction}
S=\textrm{tr}\int d^{4}x \Big(&-E^{j}E_{j}-2E^{j}\partial_{0}A_{j}-2A_{0}D_{j}(A)E^{j} \nonumber \\
&-\frac{1}{2}F_{ij}F^{ij}+2A_{\mu}J^{\mu}\Big).
\end{align}
We can now immediately integrate out $A_{0}$ since it appears as a Lagrange multiplier.  Assuming sources with fixed locations in space, $J^{j}=0, J^{0}\equiv g\rho\neq0$, then the sources will now only appear in the constraint equation
\begin{align}
Z[J]=&\int\mathcal{D}E_{j}\int\mathcal{D}A_{j}\delta(D_{j}(A)E^{j}-g\rho) \nonumber \\
&\times e^{i\textrm{tr}\int d^{4}x \big(-E^{j}E_{j}-2E^{j}\partial_{0}A_{j}-\frac{1}{2}F_{ij}F^{ij}\big)}.
\end{align}

It is now convenient to isolate the part of the chromoelectric field which is constrained from that which is not. To that end, we define the gauge-covariant-transverse and gauge-covariant-longitudinal chromoelectric fields via the relationship
\begin{equation}
E^{j}=E_{T}^{j}+E_{L}^{j},\hspace{10pt}D_{j}(A)E^{j}_{T}=0.
\end{equation}
For brevity we'll hereafter refer to these as the GC-transverse and GC-longitudinal parts. This decomposition is similar the typical transverse/longitudinal decomposition of vector fields which are defined by the relations
\begin{equation}
E^{j}=E_{L}^{j}+E_{T}^{j},\hspace{10pt}\partial_{j}E_{T}^{j}=0,
\end{equation}
and are written in terms of the transverse projection
\begin{align}
E_{L}^{j}&=\partial^{j}\big(\nabla^{-2}\partial_{k}\big)E^{k} \nonumber \\
E_{T}^{j}&=\Big(\delta^{j}_{k}-\partial^{j}\big(\nabla^{-2}\partial_{k}\big)\Big)E^{k},
\end{align}
where $\nabla^{-2}$ is the Laplace Green's function. This decomposition of the field is particularly useful because transverse and longitudinal fields are orthogonal
\begin{equation}
\int d^{3}x E_{T}^{j}E^{j}_{L}=0.
\end{equation}
This orthogonality holds also for the GC-transverse decomposition, and is key to the decomposition being useful. 

We can invert the defining relationship for these components to see that the GC-longitudinal part is a GC-gradient
\begin{equation}
E_{L}^{j}=D^{j}(A)\Big(D_{k}(A)D^{k}(A)\Big)^{-1}D_{i}(A)E^{i}.
\end{equation}
We can hereafter use a scalar (Lie-algebra valued) variable $V$, where $E_{L}^{j}=-D^{j}(A)V$. Since the second order elliptic differential operator $D_{j}(A)D^{j}(A)$ is strictly positive, its inverse is well defined~\cite{friedman1983}.

With this decomposition, the first-order form path-integral simplifies.  Firstly, the Yang-Mills Gauss law constraint does not affect $E_{T}^{j}$, and all information about the sources is contained in the exact equation
\begin{equation}\label{eq:YMgauss}
-D_{j}(A)D^{j}(A)V=g\rho,
\end{equation}
where $A_{j}$ is a background and $V$ is the variable to be solved for.

The next simplification is that the chromoelectric energy term separates because the two components are orthogonal
\begin{align}
\textrm{tr}\int d^{3}x E^{j}E_{j}&=\textrm{tr}\int d^{3}x \big(E_{L}^{j}E_{L}^{j}
+E_{T}^{j}E_{T}^{j}\big) \nonumber \\
&=\textrm{tr}\int d^{3}x \big(g\rho V[\rho,A_{j}]
+E_{T}^{j}E_{T}^{j}\big).
\end{align}
In the first term we used the constraint equation to rewrite $E_{L}^{j}$ in terms of the Yang-Mills charge density and the solution $V[\rho,A_{j}]$ to the constraint equation~\cref{eq:YMgauss}.

The final simplification comes in the ``$p\dot{q}$'' term in~\cref{eq:foaction}. It can be rewritten as
\begin{align}
2\,\textrm{tr}&\int d^{4}x E^{j}\partial_{0}A_{j}= \nonumber \\
&2\,\textrm{tr}\int d^{4}x \bigg(E_{T}^{j}\partial_{0}A_{j}-V[\rho,A_{j}]\big(D^{j}(A)\partial_{0}A_{j}\big)\bigg).
\end{align}
We can eliminate this second term by a judicious choice of gauge. Indeed, it has been proven that the so-called generalized Coulomb gauge,
\begin{equation}
D^{j}(A)\partial_{0}A_{j}=0,
\end{equation}
is a valid gauge condition which is free of Gribov ambiguities~\cite{friedman1983,cronstrom98a,cronstrom98b}. This gauge choice has a nice geometrical interpretation~\cite{friedman1983}, the details of which we will not discuss here, which make it quite useful. There has been recent work trying to use this gauge choice in the canonical quantization using the constraint formalism of Dirac, however the authors conclusions were that the canonical formulation was far more complicated than a path-integral formulation~\cite{gumus2017}.  

All together then, we have the first-order form for the generating functional
\begin{align}
Z[J]=&\int\mathcal{D}A_{j}\int\mathcal{D}E^{T}_{j}\,\Delta[A]\delta(D^{j}(A)\dot{A}_{j})e^{-i\textrm{tr}\int d^{4}x g\rho V[\rho,A_{j}]} \nonumber \\
&\times e^{i\textrm{tr}\int d^{4}x \big(-E_{T}^{j}E^{j}_{T}-2E^{j}_{T}\partial_{0}A_{j}-\frac{1}{2}F_{ij}F^{ij}\big)},
\end{align}
where $\Delta[A]$ is the Faddeev-Popov determinant corresponding to the generalized Coulomb gauge. At this stage, we can conveniently integrate out the GC-transverse electric field to obtain a Lagrangian form
\begin{align}
Z[J]=&\int\mathcal{D}A_{j}\,\Delta[A]\delta(D^{j}(A)\dot{A}_{j})e^{-i\textrm{tr}\int d^{4}x g\rho V[\rho,A_{j}]} \nonumber \\
&\times e^{i\textrm{tr}\int d^{4}x \big(\partial_{0}A_{j}\partial_{0}A^{j}-\frac{1}{2}F_{ij}F^{ij}\big)}.
\end{align}

The most important difference between the non-abelian and abelian cases is that now the solution $V[\rho,A_{j}]$ depends on the ``background'' $A_{j}$. To compute the interaction energy between static sources, we must then solve \cref{eq:YMaction} for general background $A_{j}$ and then evaluate the functional integral over $A_{j}$. 

The interaction energy between the sources is then given by the effective Hamiltonian $\mathcal{H}[\rho]$ defined by
\begin{equation}\label{eq:effhamiltonian}
e^{-i\int dt\mathcal{H}[\rho]}=\langle e^{-i\,\textrm{tr}\int d^{4}x g\rho V[\rho,A_{j}]}\rangle,
\end{equation}
where the angled brackets denote the vacuum expectation value for gauge fields in generalized Coulomb gauge,
\begin{align}
\langle\mathcal{O}[A]\rangle=&Z[0]^{-1}\int\mathcal{D}A_{j}\,\Delta[A]\delta(D^{j}(A)\dot{A}_{j})\,\mathcal{O}[A]\nonumber \\
&\times e^{i\textrm{tr}\int d^{4}x \big(\partial_{0}A_{j}\partial_{0}A^{j}-\frac{1}{2}F_{ij}F^{ij}\big)}.
\end{align}
The interaction energy between static sources has now been isolated, and it remains then to solve the linear differential equation~\cref{eq:YMgauss} and to evaluate the functional integration over backgrounds.



\section{III: Yang-Mill Gauss law}
\label{sec:YMgauss}

To understand the constraint equation we're going to expand out the Lie-Algebra valued fields in terms of the generators and work with components, $V=V^{a}T^{a},\,\rho=\rho^{a}T^{a}$, with $a=1,...,\textrm{dim(G)}$. The constraint equation then reads
\begin{align}\label{eq:gausslaw}
&-\partial_{j}\partial^{j}V^{a}+2gf^{abc}A^{b}_{j}\partial_{j}V^{c}-g^{2}f^{abc}f^{cde}A^{b}_{j}A^{d}_{j}V^{e}= \nonumber \\
&g\rho^{a}-gf^{abc}V^{c}\partial_{j}A_{j}^{b},
\end{align}
where $f^{abc}$ are the group's structure constants, $[T^{a},T^{b}]=if^{abc}T^{c}$.
We won't attempt to solve this equation for a general background, rather we're going to try and understand the nature of the solution when there is a gluon condensate.

One can make a simple intuitive argument suggesting the existence of a gluon condensate, here we borrow the argument of~\cite{fukuda1978}. Yang-Mills theory contains a three-point vertex, and thus ``H'' diagrams describing the interaction between gluons via the exchange of a gluon. It has been demonstrated that in the singlet channel the attractive force described by this gluon exchange is dominant over the repulsive force described by the four-gluon vertex. For very long-wavelength particles the binding energy will have larger magnitude than the kinetic energy of the particles and they will form a negative energy (bound) state.  Moreover, since gluons are massless the bound states will appear as tachyonic poles in correlation functions and the entire field will be unstable to a pairing condensation completely analogous to the cooper pair condensation phenomenon in BCS superconductors.  One can substantiate this intuition quantitatively by using Bethe-Salpeter equations~\cite{fukuda1977,fukuda1978,gusynin1978}. One can also argue for the instability of the empty Yang-Mills vacuum using background field methods~\cite{savvidy1977,neilsen1978}. Additionally, recent numerical-lattice evidence has emerged suggesting that the Yang-Mills vacuum may be best described by an instanton liquid, the dimension-2 condensate being a consequence~\cite{boucaud2002}.

The gluon pairing intuition suggests that while $\langle A_{j}^{a}\rangle=0$ one may still find non-zero vacuum expectation values for eg. the gauge-dependent dimension-2  operator $\langle A^{a}_{\mu}A^{a}_{\mu}\rangle$ or for the gauge-independent dimension-4 operator $\langle F^{a}_{\mu\nu}F^{a}_{\mu\nu}\rangle $. In the context of quark confinement, where one expects a dynamically generated QCD string tension $\sigma$ to arise, it is the dimension-2 condensate which has the correct dimensions to generate a tension. It isn't obvious how this should occur though because $A^{a}_{\mu}A^{a}_{\mu}$ is not a gauge invariant observable.

Despite the dimension-2 condensate being gauge dependent, one can minimize its value over the gauge group.  This minimal value is gauge invariant, and furthermore one finds that it is obtained in Landau gauge~\cite{gubarev2001}. Since lattice QCD simulations have been performed in Landau gauge which demonstrate a non-zero value for the condensate~\cite{boucaud2000a,boucaud2000b,boucaud2001}, it is expected that regardless of gauge condition one should expect a positive definite value for the dimension-2 condensate.

To describe a gluon condensate we will use a pairing parameter $\phi_{0}$, and assuming vacuum expectation values,
\begin{equation}\label{eq:pairingcond}
\langle A_{j}^{a}\rangle=0,\hspace{15pt}\langle A^{a}_{j}A^{a}_{j}\rangle=\phi^{2}_{0}.
\end{equation}
This approach has been discussed in the literature by~\cite{fukuda1982,celenza1986}. We will not aim to calculate $\phi_{0}$, rather we will investigate the consequences of it being non-zero. It is natural to expect however that $\phi_{0}$ would take a value of order the QCD scale, $\phi_{0}\sim 0.2\,\textrm{GeV}$. We will return to this point later.

To describe the gluon pairing condensate we'll follow~\cite{celenza1986} and decompose the gluon field in the path-integral as
\begin{equation}
A^{a}_{j}(x)=\mathscr{A}^{a}_{j}(x)+\phi_{0}\eta^{a}_{j},
\end{equation}
where $\mathscr{A}^{a}_{j}(x)$ has no infinite wavelength mode, $\phi_{0}$ is constant. Translation invariance will then ensure $\langle\mathscr{A}^{a}_{j}\rangle=0$. The tensor $\eta^{a}_{j}$ is constrained only to satisfy the normalization
\begin{equation}
\eta^{a}_{j}\eta^{a}_{j}=1.
\end{equation}
We'll then treat it as a $\beta_{G}$-dimensional unit vector with no preferred direction, where
\begin{equation}
\beta_{G}\equiv(d-1)\,\textrm{dim(G)},
\end{equation} 
and we'll always restrict to $d=4$ spacetime dimensions throughout. The intuitive picture of this description of the vacuum is similar to that of spontaneous symmetry breaking. The effective potential is assumed to have a non-trivial minimum which gives a vacuum expectation value of $\phi_{0}^{2}$ to the pairing field $A^{a}_{j}A^{a}_{j}$. Every semiclassical solution then has $A^{a}_{j}=\phi_{0}\eta^{a}_{j}$, however the vacuum is in a uniform superposition of equally likely \textit{condensate angles} $\eta^{a}_{j}$, so that ultimately the vacuum expectation value of the gauge field is still vanishing, $\langle A^{a}_{j}\rangle=0$.

Since we are ultimately interested in the static-long range force between sources, we will just formally evaluate the functional integration over the short-wavelength degrees of freedom $\mathscr{A}^{a}_{j}(x)$. We assume that this integration does two things, i) it generates an effective potential which allows $\phi_{0}$ to take a non-zero and rigid value, and ii) that it leads to a running coupling $g$ in our Gauss' law constraint equation~\cref{eq:gausslaw}. We will not compute this scale dependence perturbatively as usual because we will not need the form of the function $g(p^{2})$, rather we will just need certain assumptions about this function. We'll make our assumptions explicit at a later point when necessary.

With the above assumptions, the complicated Yang-Mills functional integral is reduced to a simple integral over the condensate angle $\eta^{a}_{j}$ such that vacuum expectation values are computed as
\begin{equation}\label{eq:sphereavg}
\langle\mathcal{O}[A]\rangle=\frac{\int d\eta^{a}_{j}\delta(\eta^{a}_{j}\eta^{a}_{j}-1)\,\mathcal{O}[\phi_{0}\eta^{a}_{j}]}{\int d\eta^{a}_{j}\delta(\eta^{a}_{j}\eta^{a}_{j}-1)}.
\end{equation}
In this model, correlation functions for the $\eta^{a}_{j}$ can be computed by differentiating the simple generating function
\begin{equation}
z[b^{a}_{j}]=\sum_{m=0}\frac{1}{m!}\bigg(-\frac{1}{4}b^{a}_{j}b^{a}_{j}\bigg)^{m}\frac{\Gamma(\beta_{G}/2)}{\Gamma(\beta_{G}/2+m)},
\end{equation}
where $\Gamma(x)$ is the Euler gamma function. A few examples of vacuum correlators are
\begin{align}
\langle\eta^{a}_{i}\eta^{b}_{j}\rangle&=\frac{\delta^{ab}\delta_{ij}}{\beta_{G}} \nonumber \\
\langle\eta^{a}_{i}\eta^{b}_{j}\eta^{c}_{k}\eta^{d}_{l}\rangle&=\frac{(\delta^{ab}\delta^{cd}\delta_{ij}\delta_{kl}+\textrm{all other ctrns.})}{\beta_{G}(\beta_{G}+2)} 
\end{align}
Note that this model is approximately gaussian in the sense that the higher order connected correlation functions are suppressed by factors of $\beta_{G}$. For example,
\begin{align}\label{eq:conncorr}
&\langle\eta^{a}_{i}\eta^{b}_{j}\eta^{c}_{k}\eta^{d}_{l}\rangle_{conn.} \nonumber \\
&\equiv\langle\eta^{a}_{i}\eta^{b}_{j}\eta^{c}_{k}\eta^{d}_{l}\rangle-\langle\eta^{a}_{i}\eta^{b}_{j}\rangle\langle\eta^{c}_{k}\eta^{d}_{l}\rangle-\langle\eta^{a}_{i}\eta^{c}_{k}\rangle\langle\eta^{b}_{j}\eta^{d}_{l}\rangle-\langle\eta^{a}_{i}\eta^{d}_{l}\rangle\langle\eta^{b}_{j}\eta^{c}_{k}\rangle \nonumber \\
&=-\frac{1}{\beta_{G}}\frac{(\delta^{ab}\delta^{cd}\delta_{ij}\delta_{kl}+\textrm{all other ctrns.})}{\beta_{G}(\beta_{G}+2)}.
\end{align}
In the large-$N$ limit of $SU(N)$ theory, these higher order connected correlators would than vanish at least as fast as $N^{-2}$.

With the above description of the gluon condensate, and short-wavelength modes integrated out to give a running coupling, we arrive at the following effective Gauss law constraint equation which is local in Fourier space
\begin{align}\label{eq:gausslaw2}
p^{2}V^{a}(p)+&2ig\phi_{0}f^{abc}\eta^{b}_{j}p_{j}V^{c}(p) \nonumber \\
&-(g\phi_{0})^{2}f^{abc}f^{cde}\eta^{b}_{j}\eta^{d}_{j}V^{e}(p)=
g\rho^{a}(p).
\end{align}
This is now just a set of linear algebraic equations which could in principle be solved exactly. We will not attempt this here, rather we will set up a series expansion in powers of the condensate strength $g\phi_{0}$. Since $\phi_{0}$ has dimensions of inverse length, the validity of this expansion will ultimately determined by the smallness of $r\phi_{0}$, where $r$ is a length scale characterizing the source charge density. We will soon see this explicitly.

To proceed we'll assume an expansion of the form
\begin{equation}
V^{a}(p)=\sum_{n}(g\phi_{0})^{n}V_{(n)}^{a}. 
\end{equation}
The lowest order solutions are
\begin{align}
V_{(0)}^{a}&=\frac{g\rho^{a}}{p^{2}}, \nonumber \\
V_{(1)}^{a}&=-2if^{abc}\frac{p_{j}\eta^{b}_{j}}{p^{2}}\left(\frac{g\rho^{c}}{p^{2}}\right),
\end{align}
and for $n\geq2$ we have the recursion relation
\begin{equation}\label{eq:recurrence}
V^{a}_{(n+2)}=-2if^{abc}\frac{p_{j}\eta^{b}_{j}}{p^{2}}V^{c}_{(n+1)}+f^{abc}f^{cde}\frac{\eta^{b}_{j}\eta^{d}_{j}}{p^{2}}V^{e}_{(n)}.
\end{equation}
The interaction energy (before averaging over $\eta^{a}_{j}$) then has a series expansion
\begin{equation}\label{eq:energy}
H=\frac{1}{2}\sum_{n}\int\frac{d^{3}p}{(2\pi)^{3}}(g\phi_{0})^{n}g\rho^{a}(-p)V^{a}_{(n)}(p)
\end{equation}

Some low order terms in the solution which are relevant to our upcoming discussion are
\begin{equation}\label{eq:v2}
\rho^{a}V^{a}_{(2)}=\frac{g\rho^{a}f^{a\alpha b}f^{b \beta c}\rho^{c}}{p^{4}}\eta^{\alpha}_{j}\eta^{\beta}_{k}\bigg(\delta_{jk}-4\frac{p_{j}p_{k}}{p^{2}}\bigg),
\end{equation} 
\begin{align}\label{eq:v4}
&\rho^{a}V^{a}_{(4)}=\frac{g\rho^{a}f^{a\alpha b}f^{b\beta c}f^{c\gamma d}f^{d\delta e}\rho^{e}}{p^{6}} \nonumber \\
&\times\eta^{\alpha}_{i}\eta^{\beta}_{j}\eta^{\gamma}_{k}\eta^{\delta}_{l}\bigg(16\frac{p_{i}p_{j}p_{k}p_{l}}{p^{4}}+\delta_{ij}\delta_{kl} \nonumber \\
&-4\delta_{kl}\frac{p_{i}p_{j}}{p^{2}}-4\delta_{ij}\frac{p_{k}p_{l}}{p^{2}}-4\delta_{jk}\frac{p_{i}p_{l}}{p^{2}}\bigg).
\end{align}
Note the factor of two in the recurrence relation~\cref{eq:recurrence}. This factor of two is responsible for the relative factor of four appearing in the tensor structure in~\cref{eq:v2}. Without this factor of four the tensor structure would be the standard transverse projector appearing often in QED calculations, but crucially this factor of four ensures that the longitudinal term dominates. 

In the $n^{th}$ order term in the solution there is a purely spatial (kinematic) tensor and a purely group theoretic tensor of $n$ structure constants contracted between the two source charges, and these color and kinematic structures are contracted together by the product  of $n$ $\eta^{a}_{j}$'s. Because of the rotational invariance of the vacuum state, we know that, after the spatial tensor is contracted with the functionally integrated $\eta^{a}_{j}$'s, it will be independent of $p$. We can then conclude that all of the $p$ dependence in the solution $V^{a}_{(n)}$ will be in the charge densities $\rho^{a}$, the running coupling $g(p^{2})$, and in the overall factors $p^{-2-n}$. Once we've specified the charge densities, we can then evaluate the fourier integral in ~\cref{eq:energy} without yet specifying details about the state of the internal colour degrees of freedom or the condensate angle $\eta^{a}_{j}$.

To set this up, we can implicitly define the $p$-independent part at each order, $G^{ab}_{(n)}\rho^{b}\equiv p^{-2-n}V^{a}_{(n)}$. We then have the interaction energy
\begin{equation}\label{eq:intenergy}
H=\sum_{n}\frac{1}{2}\phi_{0}^{n}G_{(n)}^{ab}\int \frac{d^{3}p}{(2\pi)^{3}}g^{2+n}(p^{2})\frac{\rho^{a}(-p)\rho^{b}(p)}{p^{2+n}},
\end{equation}
where $G^{ab}_{(0)}=\delta^{ab}$, $G^{ab}_{(2)}$ and $G^{ab}_{(4)}$ can be read off of \cref{eq:v2,eq:v4}, and higher order terms can be computed from the recurrence relation~\cref{eq:recurrence}. To evaluate this integral we'll need to specify the spatial form the of charge density, and make some assumptions about the nature of the running coupling function.

Before proceeding to understand the sources, we can further simplify this expression.  From the definition of the effective Hamiltonian,~\cref{eq:effhamiltonian}, we need to exponentiate~\cref{eq:intenergy} and evaluate the functional integral over the condensate angles $\eta^{a}_{j}$. We've observed however,~\cref{eq:conncorr}, that the higher-order connected correlation functions are suppressed by powers of $\beta_{G}^{-1}$, which is eg. for $SU(3)$ theory $\beta^{-1}_{SU(3)}=1/24$. To leading order in a $\beta_{G}^{-1}$ expansion we can then simply retain the mean field approximation for the effective Hamiltonian
\begin{equation}
\langle e^{-i\,\textrm{tr}\int d^{4}x g\rho V[\rho,A_{j}]}\rangle= e^{-i\,\textrm{tr}\int d^{4}x g\rho \langle V[\rho,A_{j}]\rangle},
\end{equation}
where the angled brackets denote the condensate angle average~\cref{eq:sphereavg}. We can then write the effective Hamiltonian for the Yang-Mills charges as
\begin{equation}\label{eq:effham}
\mathcal{H}=\sum_{n}\frac{1}{2}\phi_{0}^{n}\langle G_{(n)}^{ab}\rangle\int \frac{d^{3}p}{(2\pi)^{3}}g^{2+n}(p^{2})\frac{\rho^{a}(-p)\rho^{b}(p)}{p^{2+n}}.
\end{equation}

For later use we list a few low-order averages of the $G^{ab}_{(n)}$. Clearly $\langle G^{ab}_{(0)}\rangle=\delta^{ab}$, and at lowest non-trivial order we have
\begin{align}
\langle G^{ab}_{(2)}\rangle&=f^{a\alpha c}f^{c\beta b}\frac{\delta^{\alpha\beta}\delta_{jk}}{\beta_{G}}\left(\delta_{jk}-4\frac{p_{j}p_{k}}{p^{2}}\right) \nonumber \\
&=-\frac{f^{a\alpha c}f^{c\alpha b}}{\beta_{G}} \nonumber \\
&=\delta^{ab}\frac{C_{A}}{\beta_{G}},
\end{align}
where $C_{A}$ is the quadratic Casimir eigenvalue in the adjoint representation. At the next order we find
\begin{equation}
\langle G^{ab}_{(4)} \rangle =\frac{1}{\beta_{G}(\beta_{G}+2)}\Big(-4C_{A}^{2}\delta^{ab}+7f^{\alpha\beta\gamma}(f^{\alpha}_{\,\,ac}f^{\beta}_{\,\,cd}f^{\gamma}_{\,\,db})\Big).
\end{equation}
For $SU(N)$, there is a nice identity which allows us to see that the last term is proportional to the quadratic Casimir operator in the adjoint representation~\cite{haber2019}. As a result, we find for $SU(N)$ Yang-Mills theory
\begin{equation}
\langle G^{ab}_{(4)} \rangle =-\delta^{ab}\frac{C_{A}^{2}}{2\beta_{G}(\beta_{G}+2)}.
\end{equation}



\section{IV: Sources}
\label{sec:sources}

\subsection{Charge Density}

The formalism established thusfar is applicable to any matter source which is i) described by a gauge invariant action and ii) static ($J^{a}_{j}=0,\,J^{a}_{0}\neq0$). As a first application of this formalism we'll specify to the case of a heavy quark-antiquark pair separated by a fixed distance $r$. We use the term quark loosely to describe a particle in some representation $R$ of the gauge group, its corresponding antiquark being in the conjugate representation. The Yang-Mills charge density for this quark-antiquark pair can be written as
\begin{equation}\label{eq:chargedensitydefn}
\rho^{a}(x)=\rho^{a}_{q}\delta^{(3)}(x-r/2)+\rho^{a}_{\bar{q}}\delta^{(3)}(x+r/2).
\end{equation}
We'll soon discuss the actual form of $\rho^{a}_{q},\rho^{a}_{\bar{q}}$, but for now we'll first note that the numerator in the interaction energy is
\begin{equation}
\rho^{a}(-p)\rho^{b}(p)=\rho^{a}_{q}\rho^{b}_{q}+\rho^{a}_{\bar{q}}\rho^{b}_{\bar{q}}+\rho^{a}_{q}\rho^{b}_{\bar{q}}e^{i\vec{p}\cdot\vec{r}}+\rho^{a}_{\bar{q}}\rho^{b}_{q}e^{-i\vec{p}\cdot\vec{r}}.
\end{equation}
Clearly the first two terms are self-energy terms which are independent of the mutual separation, and we'll simply ignore these. The interaction energy can then be written as
\begin{equation}\label{eq:energyintergral}
\mathcal{H}=(\rho^{a}_{q}\rho^{b}_{\bar{q}}+\rho^{a}_{\bar{q}}\rho^{b}_{q})\sum_{n}\frac{1}{2}\phi_{0}^{n}\langle G_{(n)}^{ab}\rangle\int \frac{d^{3}p}{(2\pi)^{3}}g^{2+n}(p^{2})\frac{e^{i\vec{p}\cdot\vec{r}}}{p^{2+n}}.
\end{equation}

At the point we must make certain assumptions about the running coupling $g(p^{2})$. The full non-perturbative form of this function, valid in both the UV and IR is obviously not known. As a result of asymptotic freedom one can approximate the functional perturbatively in the UV, but there is significant disagreement about the IR form. Indeed, it is not agreed upon whether as $p^{2}\rightarrow 0$ the coupling diverges, vanishes, or freezes at a finite value (see~\cite{deur2016} for a recent review). Even within these three camps there is quantitative disagreement~\cite{deur2016}. We will not attempt to discuss all of the reasons for disagreement in the literature, but a primary issue is that different approaches lead to different definitions of the running coupling. For our purposes we need to understand whether our approach is more in-line with those suggesting $g$ diverges in the IR or those suggesting the contrary. 

It seems that generically, one finds an IR divergent running coupling if the interaction potential is written as
\begin{equation}
V(p)\sim \frac{g^{2}(p^{2})}{p^{2}},
\end{equation}
and all non-Coulombic behaviour is folded into the running coupling~\cite{deur2016}. In our expression for the interaction potential~\cref{eq:intenergy}, the running coupling is separate from a series of terms of increasing inverse powers of $p$. We then expect that the divergent behaviour as $p^{2}\rightarrow 0$ is already accounted for and will not arise in the running coupling.

 In what follows we will assume that $g(p^{2})$ reaches a non-zero freezing as $p^{2}\rightarrow 0$. We know from asymptotic freedom that the coupling vanishes as $p^{2}\rightarrow\infty$ and we'll assume that the analytic continuation also vanishes as $p^{2}$ is taken to complex infinity in the upper half complex $p$ plane. We'd further like to understand the analytic properties of $g(p^{2})$ so that we could use contour integration. By causality considerations, as a function of the four-momentum squared, $Q^{2}$, one expects observables to be analytic except along the negative real (time-like) $Q^{2}$ axis: such singularities correspond to the production of on-shell particles~\cite{gardi1998}. As a consequence, in static  $(Q^{0}=0)$ situations like the one under consideration, one might expect to have singularities along the imaginary $|\vec{p}|$ axis. However, it is not clear whether this intuition is appropriate here since our definition of the running coupling is not Lorentz invariant.  In the following contour integrals we'll neglect contributions from any singularities of $g(p^{2})$ in the complex-$|p|$ plane, assuming the singularities at $p=0$ in the integrand are dominant. In future work one certainly needs to revisit these assumptions with a more sophisticated analysis.
 
 We can then start to understand the Fourier integral in~\cref{eq:energyintergral}. For $n=0$, the fourier integral is just the Coulomb expression with running coupling
  \begin{equation}
 I_{(0)}(a)\equiv\int \frac{d^{3}p}{(2\pi)^{3}}g^{2}(p^{2})\frac{e^{i\vec{p}\cdot\vec{r}}}{p^{2}},
  \end{equation}
  and it follows from our above discussion the running coupling in this expression will not drastically change the qualitative shape of the effective Coulomb potential. After a few standard manipulations, the integral for the higher order terms can be evaluated as a residue integral
 \begin{align}
 &I_{n}(a) \nonumber \\
 &=\frac{1}{4\pi}\textrm{Re}\int^{\pi/2}_{0}d\theta\sin\theta\,i\textrm{Res}\left(\frac{g^{2+n}(p^{2})e^{ip|r\cos\theta|}}{p^{n}}\right)\bigg|_{p=0},
 \end{align}
 where it remains to evaluate the residue of the function within the parentheses. For $m>1$ we have a higher-order pole and the residue will involve various derivatives of this function. The result is
\begin{align}\label{eq:residue}
I_{n}(a)=\frac{|r|^{n-1}}{4\pi}\textrm{Re}\sum_{k=0}^{n-1}&\,i^{n-k}\frac{(n-1)!}{(n-1-k)!(n-k)!k!} \nonumber \\
&\left(\frac{1}{|r|^{k}}\frac{d^{k}}{dp^{k}}g^{2+n}(p^{2})\right)\bigg|_{p=0}.
\end{align}

If we had exact knowledge of the running coupling function, we could compute the required derivatives, but we do not have this information. We can however refer to~\cite{deur2016} wherein many plots are collected from a variety of papers suggesting IR freezing for the running coupling. It appears the running coupling is constant over a large range of values near $p=0$, and thus its derivatives are very small. The terms involving higher derivatives of the running coupling will then dominate the series in~\cref{eq:residue} only at very small separations $r$. However because these derivatives are also expected to be small relative to $\phi_{0}^{-1}$ the overall contribution to the energy from these non-Coulombic terms will be highly suppressed in this short distance limit.  To dominant order we can then keep just the $k=0$ term in~\cref{eq:residue}, and find the Hamiltonian
\begin{align}\label{eq:energyintergral2}
\mathcal{H}=&(\rho^{a}_{q}\rho^{a}_{\bar{q}})\int \frac{d^{3}p}{(2\pi)^{3}}g^{2}(p^{2})\frac{e^{i\vec{p}\cdot\vec{r}}}{p^{2}} \nonumber \\
&+(\rho^{a}_{q}\rho^{b}_{\bar{q}}+\rho^{a}_{\bar{q}}\rho^{b}_{q})\frac{g^{2}_{0}}{8\pi}\sum_{n=1}^{\infty}(-1)^{n}\langle G_{(2n)}^{ab}\rangle\frac{(\phi_{0}g_{0})^{2n}|r|^{2n-1}}{(2n)!},
\end{align}
where $g_{0}$ is the freezing value of the running coupling at $p^{2}=0$.

\subsection{Color Factors}

At this point we must work to understand the color factor, $G_{(2n)}^{ab}(\rho^{a}_{q}\rho^{b}_{\bar{q}}+\rho^{a}_{\bar{q}}\rho^{b}_{q})$, at least for the first few orders. The factor $G^{ab}_{(2n)}$ still depends on a product of $(2n)$ condensate angles $\eta^{a}_{j}$ which will need to be functionally integrated over, but also, the internal color degrees of freedom contained in $\rho^{a}_{q},\rho^{a}_{\bar{q}}$ are also quantum mechanical and this must be described as well. The effective Hamiltonian $\mathcal{H}[\rho]$ defined in~\cref{eq:effhamiltonian} is either an operator on the color Hilbert space, or equivalently, a function of color variables in a path-integral. To extract an energy which depends only on the quark-antiquark separation we'll need to average over the color degrees of freedom appropriately.

The most common discussion of the quark-antiquark pair involves the Wilson loop~\cite{kgwilson1974},
\begin{equation}
W[C]=\textrm{tr}\,\mathcal{P}\exp\left(i\oint A\right),
\end{equation}
where $\mathcal{P}$ denotes path ordering along the closed curve $C$, and the trace is performed in some representation $R$ of the gauge group. If we specify to quarks in the fundamental representation of the gauge group $SU(N)$, it is straightforward to see that this Wilson loop comes from a path-integral over color degrees of freedom~\cite{dtong2018}. Explicitly it is
\begin{equation}\label{eq:colorpathintegral}
W[C]=\int\mathcal{D}\lambda\mathcal{D}w\mathcal{D}w^{\dagger}\,e^{iS_{w}[A]}w_{j}(\tau=\tau_{f})w^{\dagger}_{j}(\tau=\tau_{i}),
\end{equation}
where $w,w^{\dagger}$ are complex $N$-dimensional vectors and the action is
\begin{equation}
S_{w}[A]=\int_{\tau_{i}}^{\tau_{f}}d\tau \bigg[iw^{\dagger}\bigg(\frac{d}{d\tau}-iA(x(\tau))\bigg)w+\lambda(w^{\dagger}w-1)\bigg],
\end{equation}
and the gauge-field 1-form is
\begin{equation}
A(x(\tau))=\frac{dx^{\mu}(\tau)}{d\tau}T^{a}A^{a}_{\mu}(x(\tau)).
\end{equation}
When the gauge field is turned off this is just a spin-coherent state path integral for the ``spin''-vector of $SU(N)$.

Following common practice, we take the curve $x^{\mu}(\tau)$ to be a rectangle in spacetime composed of a straight spacelike segments of length $r=2a$ at each of the far future and past $t\rightarrow\pm\infty$ which are connected by two straight timelike lines. With this configuration the contribution from the spacelike segments is negligible and we can rewrite the above colour path integral as the product of integrals for the quark and antiquark
\begin{align}\label{eq:singletinsertions}
W[C]=&\int\mathcal{D}\lambda\mathcal{D}w\mathcal{D}w^{\dagger}\,e^{iS_{w}}\int\mathcal{D}\bar{\lambda}\mathcal{D}\bar{w}\mathcal{D}\bar{w}^{\dagger}\,e^{iS_{\bar{w}}+i\int d^{4}x J^{\mu}A_{\mu}} \nonumber \\
&\times w_{j}(t=\infty)\bar{w}_{j}(t=\infty)w^{\dagger}_{k}(t=-\infty)\bar{w}^{\dagger}_{k}(t=-\infty).
\end{align}
Here we've pulled out the gauge interaction term, $S_{w}=S_{w}[A=0]$, and defined the current density
\begin{align}\label{eq:wilsonloopcurrentdensity}
J^{a\,\mu}(x)&=\delta^{0\mu}\delta^{(3)}(x-r/2)w^{\dagger}_{j}T_{jk}^{a}w_{k} \nonumber \\
&-\delta^{0\mu}\delta^{(3)}(x+r/2)\bar{w}_{j}^{\dagger}(T^{a}_{jk})^{*}\bar{w}_{k},
\end{align}
where the asterisk denotes complex conjugation. From this form, with the initial and final state operator insertions made explicit, we can see that this Wilson loop computes a transition amplitude between an initial singlet state for the quark-antiquark pair to a final singlet state. 

It may not be immediately clear how the operator insertions at $t=\infty$ arose in this expression. Since the Wilson loop curve is a closed loop, when it is split into quark and anti-quark integrals there is an implicit delta function constraining their color variables in the future to be equal. This final state variable is integrated over in the path integral, but one can replace this with a sum over possible operator insertions. One can check that the constraint enforced by the Lagrange multiplier will set the amplitude to vanish unless there is one insertion of $\bar{w}^{\dagger}_{j}w^{\dagger}_{k}$ in the past and one insertion of $\bar{w}^{\dagger}_{l}w^{\dagger}_{m}$ in the future. The resolution of the identity in the future can then be expanded in a Fock basis and the constraint ensures that only the final singlet state leads to a non-zero result.

We can take the Wilson loop current density~\cref{eq:wilsonloopcurrentdensity} and extract from it the $SU(N)$ charges $\rho^{a}_{q},\,\rho_{\bar{q}}^{a}$ defined in~\cref{eq:chargedensitydefn},
\begin{align}\label{eq:wTw}
&\rho^{a}_{q}=w^{\dagger}_{j}T^{a}_{jk}w_{k}, \nonumber \\
&\rho^{a}_{\bar{q}}=-\bar{w}^{\dagger}_{j}(T^{a}_{jk})^{*}\bar{w}_{k}.
\end{align}
In principle, we could then compute the Yang-Mills vacuum expectation value of the Wilson-loop by evaluating the color path-integrals
\begin{align}
W[C]=&\int\mathcal{D}\lambda\mathcal{D}w\mathcal{D}w^{\dagger}\,e^{iS_{w}}\int\mathcal{D}\bar{\lambda}\mathcal{D}\bar{w}\mathcal{D}\bar{w}^{\dagger}\,e^{iS_{\bar{w}}}\,e^{-i\int dt\mathcal{H}} \nonumber \\
&\times w_{j}(t=\infty)\bar{w}_{j}(t=\infty)w^{\dagger}_{k}(t=-\infty)\bar{w}^{\dagger}_{k}(t=-\infty),
\end{align}
with $\mathcal{H}$ given by~\cref{eq:energyintergral2} and the charges written in terms of the $w_{j}$ as in~\cref{eq:wTw}. However as we will see shortly, the operator method is more straightforward.

Although we only have the explicit path-integral representation for the internal color variables of quarks in the fundamental representation of $SU(N)$, we can proceed more generally following the above statement that Wilson loops compute transition amplitudes for quark-antiquark pairs evolving from initial singlet states to final singlet states.  In an operator representation,  the Yang-Mills charges $\rho^{a}_{q},\,\rho_{\bar{q}}^{a}$ for a general representation $R$ of group $G$ would then have the form
\begin{align}\label{eq:charges}
&\rho^{a}_{q}=\delta_{\bar{A}\bar{B}}T^{a}_{AB}(R), \nonumber \\
&\rho^{a}_{\bar{q}}=-\delta_{AB}(T^{a}(R)^{T})_{\bar{A}\bar{B}},
\end{align}
where $T$ denotes the transpose, $A,B$ are indicies in the quark-color Hilbert space and $\bar{A},\bar{B}$ are indicies in the antiquark-color Hilbert space. In this basis the singlet state has wavefunction
\begin{equation}
\psi^{sing}_{A\bar{A}}=(\textrm{dim(R)})^{-1/2}\delta_{A\bar{A}}.
\end{equation}

In the Hamiltonian~\cref{eq:energyintergral2}, the color charge operators are contracted at each order with the factor $\langle G^{ab}_{(2n)}\rangle$.  We've also demonstrated that the lowest order contributions, $\langle G^{ab}_{(0)}\rangle, \langle G^{ab}_{(2)}\rangle, \langle G^{ab}_{(4)}\rangle$, are all proportional  to $\delta^{ab}$. It then follows that up to $\mathcal{O}(\phi_{0}^{4})$ the singlet state is an eigenstate of the Hamiltonian. We see this from the definition of the quadratic Casimir operator
\begin{align}
(\rho^{a}_{q}\rho^{a}_{\bar{q}})_{A\bar{A},B\bar{B}}\,\psi^{sing}_{B\bar{B}}&=
-T^{a}_{AB}(R)T^{a}(R)_{B\bar{A}}(\textrm{dim(R)})^{-1/2} \nonumber \\
&=-C_{R}\,\psi^{sing}_{A\bar{A}}.
\end{align}

When computing Wilson loops, since the color states start in an eigenstate of the Hamiltonian (up to $\mathcal{O}(\phi_{0}^{4})$), they remain in this state and we can replace the charge operator $\rho^{a}_{q}\rho^{a}_{\bar{q}}$ by the eigenvalue in the singlet state, $C_{R}$. The energy eigenvalue for a singlet state is
\begin{align}\label{eq:energyintergral4}
\mathcal{H}=&-C_{R}\int \frac{d^{3}p}{(2\pi)^{3}}g^{2}(p^{2})\frac{e^{i\vec{p}\cdot\vec{r}}}{p^{2}} \nonumber \\
&+\frac{g^{2}_{0}}{4\pi}\frac{C_{R}C_{A}}{2\beta_{G}}(\phi_{0}g_{0})^{2}|r| \nonumber \\
&+\frac{g^{2}_{0}}{4\pi} \frac{C_{R}C_{A}^{2}}{4\beta_{G}(\beta_{G}+2)}\frac{(\phi_{0}g_{0})^{4}|r|^{3}}{4!}+...\,.
\end{align}
The first two terms are universal, whereas the $r^{3}$ term has been proven here only for the $SU(N)$ gauge group. The overall proportionality of the energy to the quadratic Casimir eigenvalue of the representation is a nice check of our result thusfar. It has been demonstrated quite convincingly in lattice simulations that this ought to occur~\cite{bali1999,bali2000,shevchenko2000}. 

Assuming $r(g_{0}\phi_{0})$ is sufficiently small that we can neglect the non-linear terms in this series, we can read off an effective string tension from~\cref{eq:energyintergral4},
\begin{equation}
\sigma=\frac{g^{2}_{0}}{4\pi}\frac{C_{R}C_{A}}{2\beta_{G}}(\phi_{0}g_{0})^{2},
\end{equation}
and for fundamental quarks in $SU(N)$ theory this is
\begin{equation}\label{eq:suntension}
\sigma=\frac{g_{0}^{2}}{48\pi}(g_{0}\phi_{0})^{2}.
\end{equation}
The freezing value of the coupling, $g_{0}$, and the condensate strength $g_{0}\phi_{0}$ remain as unfixed parameters.

\section{V: Discussion}
\label{sec:disc}


There are a number of ways that we could estimate the parameters $g_{0}$ and $g_{0}\phi_{0}$ by comparing with reported values in the literature.  Unfortunately, the dimension-2 condensate strength is gauge dependent and the definition of the running coupling constant is gauge and renormalization scheme dependent. This precludes any direct comparison since we could not find previously reported results involving these quantities in the generalized Coulomb gauge. We'll instead make rough estimates based on reported values in other gauges and also try to use phenomenological constraints.

Let's first return to the assumption that $r(g_{0}\phi_{0})$ is small. We can look at the relative size of the cubic term to the linear term in the binding energy as a function of $r$.  If the magnitude of cubic term is to be less than $5\%$ of the magnitude of the linear term then we must have
\begin{equation}
r<\left(\frac{1}{g_{0}\phi_{0}}\right)\sqrt{\frac{12(\beta_{G}+2)}{5C_{A}}}.
\end{equation}
For $SU(3)$ theory, with appropriate factors of $\hbar$ and $c$ replaced, this is
\begin{equation}
r<\frac{0.9\,\textrm{fm\,GeV}}{g_{0}\phi_{0}}.
\end{equation}

One expects that at sufficiently large separations, when it is energetically favourable, the QCD string will snap and a quark-antiquark pair will be produced. This leads to a flattened static potential above some critical separation $r_{c}$. Such an effect could not be seen in our heavy quark model, but it has been observed in lattice simulations with dynamical quarks. In these simulations a value $r_{c}\approx 1.2\,\textrm{fm}$ is found consistently~\cite{bali2000b,bali2005}. 

In the ``quenched'' simulations which do not include dynamical quarks, the linear rise of the potential has been shown to continue past the $1.2\,\textrm{fm}$ mark and no cubic behavior has been conclusively demonstrated to arise~\cite{bali1992,bali2000c,bali2000d,bolder2001,necco2001}.  Furthermore, one can prove that invariance of Wilson loops under space/time interchange implies that the potential cannot grow faster that linearly with $|r|$ at large distances~\cite{seiler1978}. These considerations then cast doubt on the validity of the model here discussed. 

Despite the inconsistencies between this model and the constraints from quenched lattice data we can still ask whether the model may be rendered consistent by the inclusion of dynamical quarks, ie. string-breaking effects. Since the string-breaking phenomenon has been demonstrated to flatten the potential above $r=1.2\,\textrm{fm}$ we can require only that our non-linear terms are negligible up to this distance. This then implies the constraint
\begin{equation}
g_{0}\phi_{0}\lesssim 0.75\,\textrm{GeV}.
\end{equation}
Additionally, from a recent review, ref. \cite{deur2016}, there is a table of reported values for the freezing value of  $a_{s}(0)=g_{0}^{2}/(4\pi)$. Most of the reported values apply to calculations in Landau gauge and no such calculation has been done yet in the generalized Coulomb gauge we've chosen here. 

To get a sense of the viability of our model, we take two representative values from this table, $a_{s}(0)=2.97$ and $a_{s}(0)=4.74$. The first value is commonly found in Landau gauge calculations~\cite{deur2016,kellermann2008} whereas the second has been computed in the less frequently used Coulomb gauge~\cite{schleifenbaum2006}. We expect that a genuine analysis of the freezing value of the coupling constant in the generalized Coulomb gauge would more closely match the Coulomb gauge result than the Landau result.  We can perform a very crude estimate of the freezing coupling in the generalized Coulomb gauge by taking an average of the values from these other gauges. Taking this together with the above constraint of the condensate strength, we can estimate a constraint on our model's prediction of the QCD string tension
\begin{equation}
\sigma\lesssim \,(0.181\pm 0.024)\,\textrm{GeV}^{2}.
\end{equation}
This upper limit is consistent with the commonly understood value for the QCD string tension $\sigma\approx (0.18-0.22)\,\textrm{GeV}$~\cite{balireview}.

 Although the estimates of the free parameters $g_{0},\,g_{0}\phi_{0}$ are quite crude, it is encouraging that the string tension predicted by our calculation is not ruled out by meson phenomenology. It remains however to try and predict the value of these parameters in the generalized Coulomb gauge using tools which have been previously used in Landau or Coulomb gauge to understand the IR limit of Yang-Mills theory.  For the freezing value of the coupling, one may try to use Schwinger-Dyson techniques as in, for example,~\cite{schleifenbaum2006} (see also the review~\cite{deur2016} and refs. therein). For the condensate strength, one may try to use Bethe-Salpeter equation techniques~\cite{fukuda1977,gusynin1978,fukuda1978,fukuda1982}.

In addition to the above constraints which we've used to assess the validity of this model, we can also try to compare with calculations of the QCD string tension coming from rather different approaches. Most studies involving a dimension-2 condensate are interested in the Lorentz invariant condensate $g^{2}\langle A^{a}_{\mu}A_{a}^{\mu}\rangle$, not just the spatial parts which we've isolated. We can try to find a comparison by using the Lorentz invariance of the vacuum, which implies
\begin{equation}
\langle A^{a}_{j}A^{a}_{j}\rangle=3\langle A^{a}_{0}A^{a}_{0}\rangle.
\end{equation}
suggesting that in the commonly used Euclidean spacetime 
\begin{equation}
\langle A^{a}_{\mu}A_{a}^{\mu}\rangle = (4/3)\phi_{0}^{2}.
\end{equation}
One must be careful when comparing the RHS of this expression to our~\cref{eq:pairingcond} though; the dimension-2 condensate is in principle a gauge-dependent quantity and the above considerations may only apply in a covariant gauge. Nonetheless we can still try make contact between our calculation on the QCD string tension and some previously reported calculations which used quite different approaches. 

In the literature, investigations of the dimension-2 condensate and its studies have predominantly used  operator product expansion (OPE) techniques~\cite{boucaud2000a,boucaud2000b,boucaud2001,destoto2001,boucaud2002,OPEcondensate}, we hope that the rather different approach we've provided here may prove to be complementary.
 For example, ref.~\cite{arriola2006} discusses the connection between the QCD string tension, the tachyonic gluon mass, and the dimension-2 condensate. Their analysis is an extension of refs.~\cite{chetyrkin1999,narison2001} in which the physics of the condensate of modeled by assuming the existence of a term $~\lambda^{2}/Q^{2}$ in the OPE of various QCD correlation functions $\Pi_{J}(Q^{2})$. 

 We can actually make direct contact with some of this work. In ref.~\cite{arriola2006} the authors claim a short distance string tension for $SU(N)$ theory, 
\begin{equation}
\sigma_{0}=\frac{g^{2}}{72\pi}\frac{N^{2}}{N^{2}-1}g^{2}\langle A_{\mu}A^{\mu}\rangle,
\end{equation}
and although it isn't explicitly stated, from their references it appears they are working in Euclidean spacetime.  Using the $4/3$ factor we then translate their expression into our notation
\begin{equation}
\sigma_{0}=\frac{g_{0}^{2}}{54\pi}\frac{N^{2}}{N^{2}-1}g_{0}^{2}\phi_{0}^{2}.
\end{equation}
Although this doesn't explicitly agree with our formula~\cref{eq:suntension} for general $N$, we do find agreement for the phenomenologically interesting case of $N=3$. The disagreement for general $N$ may just be the result of their relative rescaling of the condensate for general $N$, but it is not clear. Either way it is encouraging that using either OPE techniques, or our current formalism, one can arrive at apparently identical expressions for the QCD string tension. This provides optimism that despite the limitations of this approach, it may still prove to be a useful avenue for performing further Yang-Mills calculations.


\section{VI: Conclusions}
In this paper we have provided an analysis of the static quark-antiquark binding energy in the presence of a gluon condensate. We used the first-order path-integral formalism, wherein the chromoelectric field is an explicit variable. The benefit of this approach was that the Yang-Mills constraint equation was explicit and exact. We used a gauge-covariant generalization of the typical transverse/longitudinal decomposition from electrodynamics to separate out the constrained variables from the unconstrained variables. Doing all of this, and fixing the generalized Coulomb gauge condition, we demonstrated that the static quark-antiquark binding energy is determined solely by the solution to the Yang-Mills Gauss law equation. 

We modeled the Yang-Mills vacuum as a dimension-2 gluon pairing condensate and used this to set-up a series expansion for the solution to the Yang-Mills Gauss law equation. As a consequence, we arrived at a series expansion for the static  quark-antiquark binding energy in powers of the condensate strength, with a recursion relation to compute the higher-order terms. The central result of the paper is a prediction of the coefficient of the term linear in quark separation, ie. the string tension. Our result matches a prediction coming from OPE analysis, suggesting that the techniques used here may indeed provide a useful complementary approach for more detailed calculations.

\section{Acknowlegements}
This work was funded by the National Science and Engineering Research Council of Canada (NSERC). We thank P.C.E. Stamp and C. DeLisle (UBC) for useful and related discussions.


\color{black}

\end{document}